\documentclass[11pt]{article}
\usepackage{BSMvietnam,epsfig}
\usepackage{ifthen} 
\usepackage{graphicx}  
\usepackage{xspace} 

 \def\PK      {\ensuremath{\mathrm{K}}\xspace}

 \def\Pi      {\ensuremath{\mathrm{i}}\xspace}

 \mathchardef\PDelta="7101
 \mathchardef\PXi="7104
 \mathchardef\PLambda="7103
 \mathchardef\PSigma="7106
 \mathchardef\POmega="710A
 \mathchardef\PUpsilon="7107

 \def\PK      {\ensuremath{K}\xspace}

 \def\Pi      {\ensuremath{i}\xspace}

\newcommand\TTstrut{\rule{0pt}{3.2ex}}

\newcommand\BBstrut{\rule[-1.8ex]{0pt}{0pt}}

\newcommand{\Dmm}{$D^{0} \to \mu^+ \mu^-$}

\newcommand{\bstommmma}{\Bsmumumumu}
\newcommand{\bdtommmma}{\Bdmumumumu}

\def\kaon  {\ensuremath{\PK}\xspace}

\def\Kp    {\ensuremath{\kaon^+}\xspace}

\def\Kstarz  {\ensuremath{\kaon^{*0}}\xspace}

\def\Kstar   {\ensuremath{\kaon^*}\xspace}

\newcommand{\tev}{\ensuremath{\mathrm{\,Te\kern -0.1em V}}\xspace}
\newcommand{\gev}{\ensuremath{\mathrm{\,Ge\kern -0.1em V}}\xspace}
\newcommand{\mev}{\ensuremath{\mathrm{\,Me\kern -0.1em V}}\xspace}
\newcommand{\kev}{\ensuremath{\mathrm{\,ke\kern -0.1em V}}\xspace}
\newcommand{\ev}{\ensuremath{\mathrm{\,e\kern -0.1em V}}\xspace}
\newcommand{\gevc}{\ensuremath{{\mathrm{\,Ge\kern -0.1em V\!/}c}}\xspace}
\newcommand{\mevc}{\ensuremath{{\mathrm{\,Me\kern -0.1em V\!/}c}}\xspace}
\newcommand{\gevcc}{\ensuremath{{\mathrm{\,Ge\kern -0.1em V\!/}c^2}}\xspace}
\newcommand{\gevgevcccc}{\ensuremath{{\mathrm{\,Ge\kern -0.1em V^2\!/}c^4}}\xspace}
\newcommand{\mevcc}{\ensuremath{{\mathrm{\,Me\kern -0.1em V\!/}c^2}}\xspace}

\newcommand{\Bmumu}{\Bqmumu}

\newcommand{\BsJpsiPhiMuons}{\decay{\Bs}{\Jpsi(\mumu)\Phi(\mumu)}}

\newcommand{\CL}{C.L.\ }

\newcommand{\CLs}{\ensuremath{\textrm{CL}_{\textrm{s}}}}

\newcommand{\comment}[1]{\par\noindent {\em\small [#1]}}
\renewcommand{\comment}[1]{}

\newcommand{\unit}[1]{\ensuremath{\rm\,#1}}

\newcommand{\invfb}{\unit{fb^{-1}}}

\newcommand{\BRof}[1]{\ensuremath{{\cal B}(#1)}}
\newcommand{\CP}{\ensuremath{\rm CP}}

\newcommand{\particle}[1]{{\ensuremath{\rm #1}}}

\newcommand{\Bd}{\particle{B^0}}
\newcommand{\Bs}{\particle{B^0_s}}

\newcommand{\Lb}{\particle{\Lambda_b}}

\newcommand{\Bq}{\particle{B^0_{(s)}}}

\newcommand{\Jpsi}{\particle{J\!/\!\psi}}

\newcommand{\ee}{\particle{e^+e^-}}

\newcommand{\KS}{\particle{K^0_S}}

\newcommand{\pip}{\particle{\pi^+}}
\newcommand{\pim}{\particle{\pi^-}}

\newcommand{\decay}[2]{\particle{#1\!\to #2}}

\newcommand{\mumu}{\particle{\mu^+\mu^-}}

\newcommand{\Bsmm}{\decay{\Bs}{\mu^+\mu^-}}              
\newcommand{\Bdmm}{\decay{\Bd}{\mu^+\mu^-}}              

\newcommand{\Bsmumu}{\decay{\Bs}{\mu^+\mu^-}}            
\newcommand{\Bqmumu}{\decay{\Bq}{\mu^+\mu^-}}            
\newcommand{\Bdmumu}{\decay{\Bd}{\mu^+\mu^-}}            

\newcommand{\Bsmumumumu}{\decay{\Bs}{\mu^+\mu^-\mu^+\mu^-}}            
\newcommand{\Bdmumumumu}{\decay{\Bd}{\mu^+\mu^-\mu^+\mu^-}}            

\newcommand{\beq}{\begin{equation}}
\newcommand{\eeq}{\end{equation}}

\def\ellell     {\ensuremath{\ell^+ \ell^-}\xspace}
\def\lhcb {\mbox{LHCb}\xspace}

\def\be{\begin{equation}}
\def\ee{\end{equation}}
\def\bea{\begin{eqnarray}}
\def\eea{\end{eqnarray}}

\begin{document}
\vspace*{4cm}
\title{RARE DECAYS WITH LHCB}

\author{G. MANCINELLI}

\address{ on behalf of the LHCb Collaboration\\
 Centre de Physique des Particules de Marseille (CPPM)\\
Aix-Marseille Universit\'e, France}

\maketitle\abstracts{Rare decays  involving leptons or photons in the final states are studied using 1.0\,fb$^{-1}$ of $pp$ collisions at a centre-of-mass energy of $\sqrt{s}=7$\,TeV collected by the LHCb experiment~\cite{lhcb} in 2011. We present results of measurements of branching ratios, angular distributions, and isospin asymmetries obtained using this data sample.}

\section{Introduction}

As Flavor Changing Neutral Currents (FCNC) processes are prohibited at tree level in the Standard Model (SM), they are particularly interesting as new particles entering the penguins or box diagrams can introduce measurable effects in Branching Ratios (BR), angular distributions, and asymmetries. 

For BR measurements the strategies are all quite similar. The criteria used to select $B$ and $D$ mesons are highly displaced secondary vertices, thanks to the large LHCb boost, the track and decay vertex quality, as well as geometrical and kinematic constraints. To efficiently remove backgrounds, LHCb exploits its good muon identification and its high trigger efficiency for muons down to low transverse momenta ($\sim$0.5 GeV). The yield for the mode under study is normalized to that of control channels with similar trigger and/or geometrical features in order to limit systematics uncertainties and, when necessary, the LHCb measured value of $f_s/f_d$ is used~\cite{LHCb-CONF-2011-034}.
Control samples from real data are used to avoid as much as possible the use of simulation (to calculate efficiencies and calibrate multivariate classifiers, for example). Finally all analyses are performed blindly in order to avoid unconscious biases.  The blinding process consists in hiding events for which the candidate has an invariant mass close to the nominal one. The CLs method is used to extract limits in absence of signal, otherwise extended maximum likelihood fits are usually performed to infer the yields.

\section{Branching fraction measurements of $B^{0}_{(s)} \to \mu^+ \mu^-$}\label{bsdmumu}

The dominant contribution to these modes in the SM stems from the 
Z-penguin diagram, 
while the box diagram is suppressed by a factor of $|m_W/m_t|^2$. 
The Higgs annihilation diagram contributes only negligibly (about 1/1000).
These are FCNC modes which, in addition, are also helicity suppressed. Hence the SM expectations are only~\cite{buras} $\mathcal{B}$(\Bsmm)$= (3.2 \pm 0.2)\times 10^{-9}$ and $\mathcal{B}$(\Bdmm)$= (0.10 \pm 0.01)\times 10^{-9}$.
Taking into account the oscillation of the $B^{0}_{s}$ system, its time integrated branching fraction is evaluated to be~\cite{bruyn}
$\mathcal{B}$(\Bsmm)$= 3.4 \times 10^{-9}$.
These modes are very sensitive to NP with new scalar or pseudoscalar interactions, as well as models with an extended higgs sector and high $\tan\beta$~\cite{babu}.

Each B candidate is given a likelihood to be signal or background-like in a two-dimensional space formed by the invariant mass and a multivariate classifiers, a boosted decision tree (BDT), using as input nine  variables describing event topology/kinematics, trained using simulation and calibrated with control samples from data.

\begin{figure}[htb]
  \begin{center}
    \includegraphics*[width=0.45\textwidth]{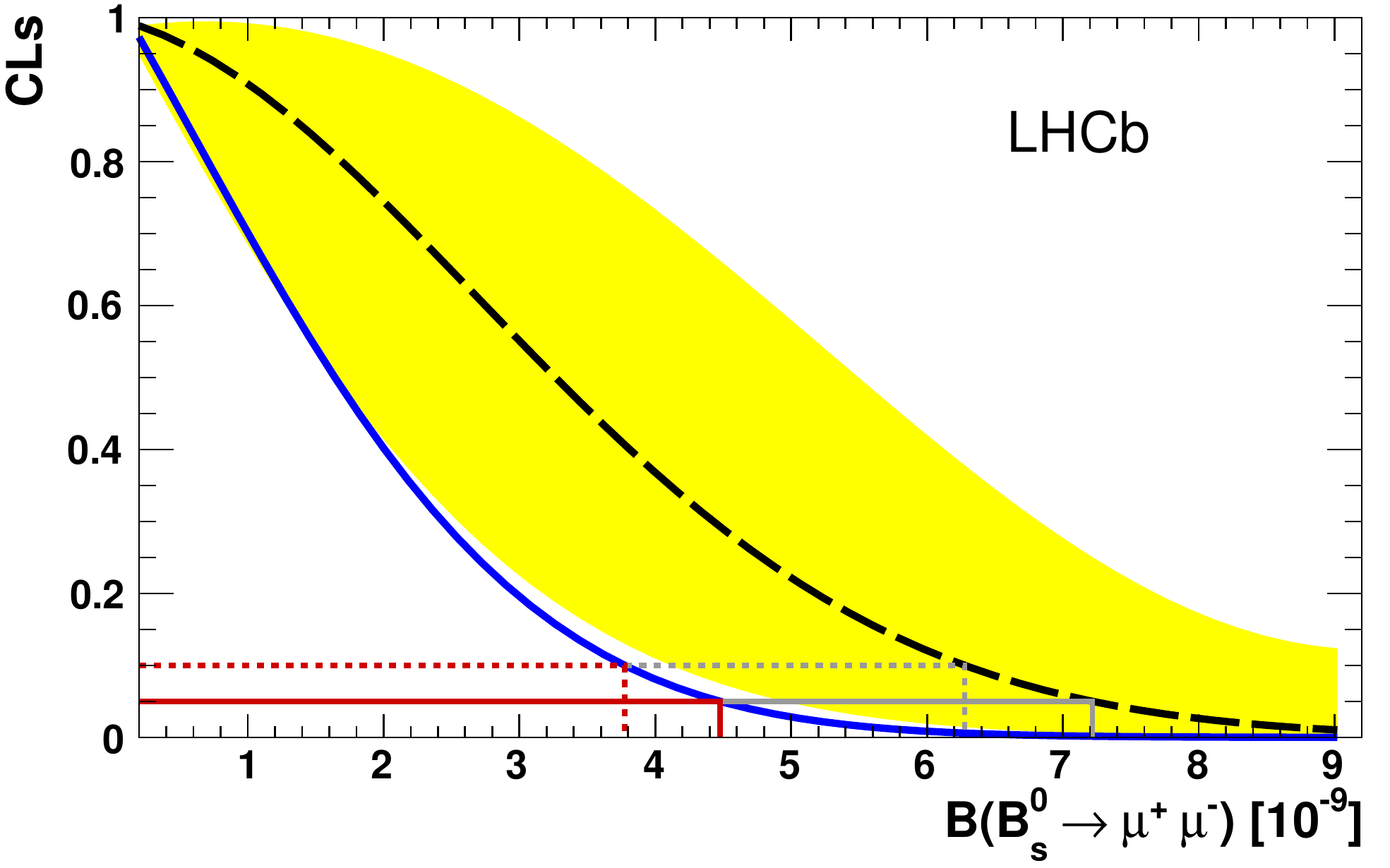}
    \includegraphics*[width=0.45\textwidth]{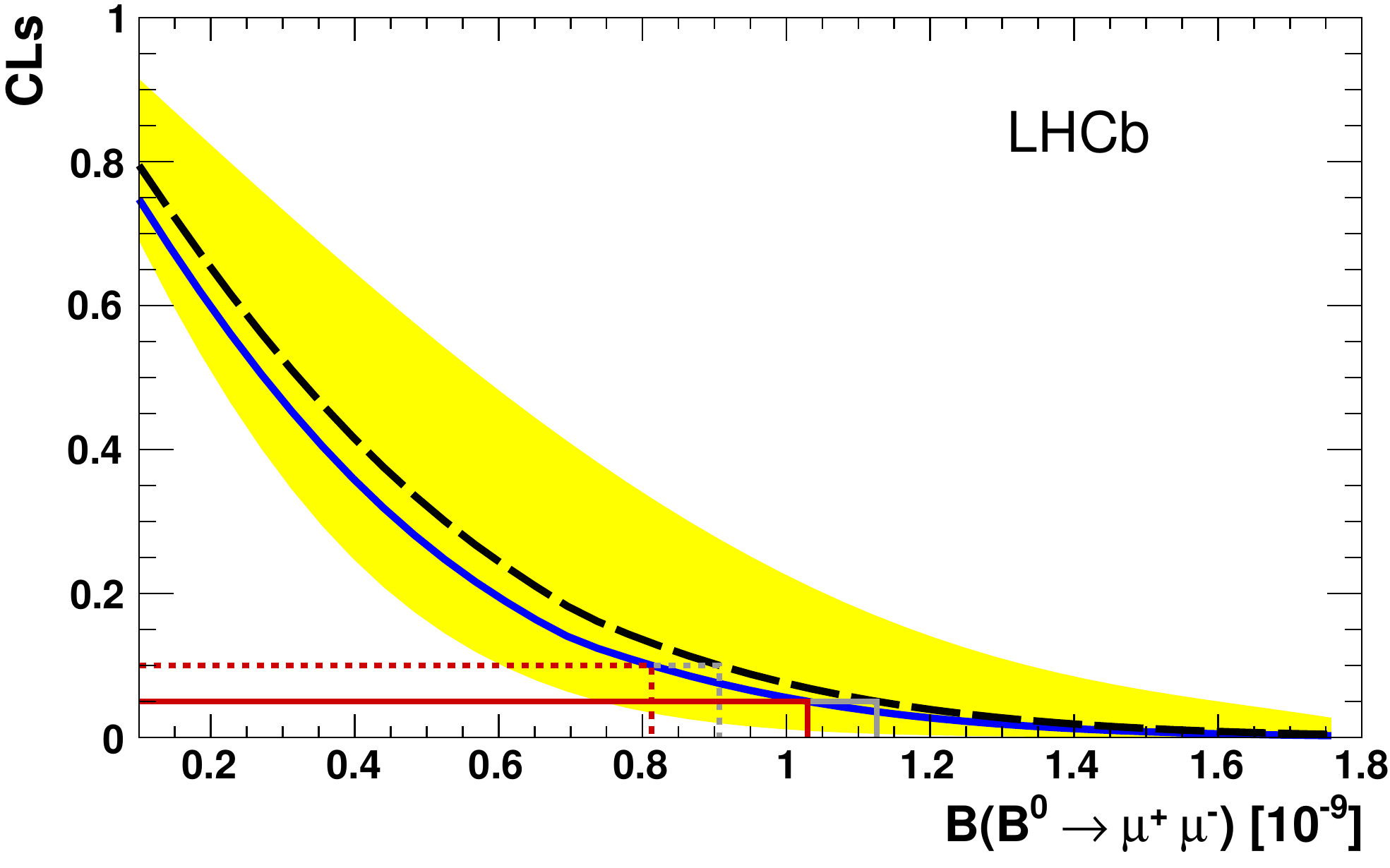}
  \end{center}
\caption{\CLs\ as a function of the assumed BR for (left) \Bsmumu and (right) 
\Bdmumu\ decays.
The long dashed black curves are the medians of the expected \CLs\ distributions for \Bsmumu, if background and SM signal were observed, and for \Bdmumu, if background only was observed.
The yellow  areas cover, for each BR, 34\% of the expected \CLs\ distribution on each side of its median.
The solid blue curves are the observed \CLs. The upper limits at 90\,\% (95\,\%) \CL are indicated by the dotted (solid) horizontal lines in red (dark gray) for the observation and in gray for the expectation.}
\label{fig:cls_bsbd}
\end{figure}

The expected and observed \CLs\ values are shown in Fig.~\ref{fig:cls_bsbd} for the \Bsmumu and \Bdmumu channels,
each as a function of the assumed branching fraction.
The expected and measured limits for \Bsmumu and \Bdmumu at 90\,\% 
and 95\,\% \CL\ are shown in Table~\ref{tab:bds_results}.

\begin{table}[!htb]
\caption{Expected and observed limits on the \Bmumu\ branching fractions, including systematics.}
\label{tab:bds_results}
\begin{center}
\begin{tabular}{@{}llcc@{}}
\hline\hline 
         Mode & Limit & at 90\,\% \CL & at 95\,\% \CL\TTstrut\BBstrut\\ 
\hline 
\Bsmumu      & Exp. bkg+SM           &  $6.3 \times 10^{-9} $  & $ 7.2  \times 10^{-9} $\TTstrut\\ 
             & Exp. bkg              &  $2.8 \times 10^{-9} $  & $ 3.4  \times 10^{-9} $   \\ 
             & Observed              &  $3.8 \times 10^{-9} $  & $ 4.5  \times 10^{-9} $   \\ 
\hline
\Bdmumu      & Exp. bkg              &  $0.91 \times 10^{-9}$  & $ 1.1 \times 10^{-9}$\TTstrut\\ 
             & Observed              &  $0.81 \times 10^{-9}$  & $ 1.0 \times 10^{-9}$   \\ 
\hline \hline
\end{tabular}
\end{center}
\end{table}

The LHCb results constitute the most constraining limits~\cite{bsmm} on these branching fractions obtained with a single experiment to date.

Recently the results of LHCb, CMS, and ATLAS have been combined for an even (slightly) more stringent limit, $\mathcal{B}(B^{0}_{s} \to \mu^+ \mu^-) < 4.2 \times 10^{-9}$, which is dominated by the LHCb contribution and is attending the SM prediction.
These results pose stringent constraints on physics beyond the SM.

\section{Branching fraction measurement of $D^{0} \to \mu^+ \mu^-$}\label{d0mumu}

The suppression of FCNC in the charm sector is driven by the GIM mechanism. The \Dmm\ decay is very rare in the SM~\cite{burdman}: $10^{-13}<\mathcal{B}(D^{0} \to \mu^+ \mu^-) < 6\times 10^{-11}$, still far from the reach of LHCb. In the context of MSSM scenarios with R parity violation, the predicted branching fractions can be largely enhanced~\cite{dmmbsm}. The best limit before LHCb was from Belle ($1.4 \times 10^{-7}$ at 90\% \CL). A two-dimensional fit in the $D^0$ mass and the $D^*-D^0$ mass difference~\cite{dmmlhcb} gives $\mathcal{B}(D^{0} \to \mu^+ \mu^-) < 1.3\times 10^{-8}$ at 95$\%$ C.L., which is the world best limit to date.

\section{Branching fraction measurement of $\tau^- \to \mu^- \mu^+ \mu^-$}\label{taumumumu}

Lepton Flavor Violation (LFV) is established in the SM, given the neutrino oscillation. Nonetheless it is an extremely suppressed phenomenon and way beyond current experimental sensitivities in the charged sector.
On the other hand charged LFV can be quite enhanced in several NP scenarios, more in $\tau$ than in $\mu$ decays.  For instance, in the context of Little Higgs models~\cite{lhmtaummm} $\mathcal{B}(\tau^- \to \mu^- \mu^+ \mu^-)<10^{-7}$. The current limits from Belle and BaBar are on the verge of being interesting and it is important to test whether this analysis can be pursued in a hadronic environment, given the large $\tau$ production cross section at the LHC.
After a loose selection, events are classified in a three-dimensional space formed by the three muons system invariant mass and two multivariate operators, one for PID information and the other for geometrical and kinematical variables.
The LHCb measurement of this branching fraction ($\mathcal{B}(\tau^- \to \mu^- \mu^+ \mu^-)<7.8 \times 10^{-8}$ at 95$\%$ C.L.)~\cite{taummmlhcb} is comparable with the current best limits.

\section{Branching fraction measurements of $B^{0}_{(s)} \to \mu^+ \mu^- \mu^+ \mu^-$}\label{bsd4mu}

The decay of a $B^{0}_{(s)}$ meson to a four muons final state is strongly suppressed in the SM and mainly occurs via the resonant mode \BsJpsiPhiMuons. In the SM, the BR of the non-resonant modes is predicted not to exceed $10^{-10}$, but could receive an enhancement due to scenarios with new particles decaying into two muons.

A standard cut and count analysis is used, optimised by means of the resonant \BsJpsiPhiMuons\ mode, whose observed yield is compatible with SM expectations, while in the non resonant no signal is found. Hence  the first limits on these processes are set to  $\BRof{\bstommmma} < 1.3\times 10^{-8}$ and $\BRof{\bdtommmma} < 5.4\times 10^{-9}$ at 95\% \CL

\section{Branching fraction measurement of $B^{+} \to \pi^{+} \mu^+ \mu^-$}\label{bpimumu}

$B^{+} \to \pi^{+} \mu^+ \mu^-$ decays constitute the first $b \to d \,\,l^+ l^-$ transitions ever observed. 
In the SM, such transitions are suppressed by loop and the CKM  factor $|V_{td}/V_{ts}|$. This is not  necessarily the case for NP models.

LHCb observes $25.3\,^{+6.7}_{-6.4}$ signal candidates, as shown in Fig.\ref{fig:angular}~\cite{pipluslhcb}, for a significance of over 5 sigma. This corresponds to  $\mathcal{B}$($B^{+} \to \pi^{+} \mu^+ \mu^-$)$= (2.4 \pm 0.6_{(stat)} \pm 0.2_{(syst)})\times 10^{-8}$, which is in agreement with the SM prediction of $\mathcal{B}$($B^{+} \to \pi^{+} \mu^+ \mu^-$)$= (1.91 \pm 0.21)\times 10^{-8}$~\cite{bpluspiplusmm}, and with a lack of large NP contribution to the  $b \to d\ellell$ processes. $B^{+} \to \pi^{+} \mu^+ \mu^-$ is the rarest $B$ decay ever observed.
The $b \to d$ transitions can show potentially larger \CP\ and isospin violating effects than their $b \to s$ counterparts due to the different CKM hierarchy~\cite{Beneke:2004dp}. These studies would need the large statistics provided by the future \lhcb Upgrade. 

\section{Angular analysis of $B^{0} \to K^{*0} \mu^+ \mu^-$}\label{bkstarmumu}

The angular distribution of the $B^{0} \to K^{*0} \mu^+ \mu^-$ decay is sensitive to the magnetic and vector and axial operators C7, C9, and C10. NP can effect composite variables and asymmetries, like the Forward Backward Asymmetry (A$_{\rm FB}$) in the $\mu\mu$ rest frame. The LHCb analysis is performed blindly, using several control samples to limit its dependence on MC in terms of trigger corrections, selection, reconstruction efficiencies, and acceptance. The selection is based on BDTs. The yield is of 900 events, more than BaBar, Belle, and CDF combined.
The angular analysis is based on the fact that the decay can be described as a function of three angles and the dimuon invariant mass and parametrized in terms of several angular observables, A$_{\rm FB}$, F$_{\rm L}$, S$_{\rm 3}$, and A$_{\rm IM}$~\cite{angular}. 

\begin{figure}[htb]
\begin{center}
    \includegraphics*[width=0.33\textwidth]{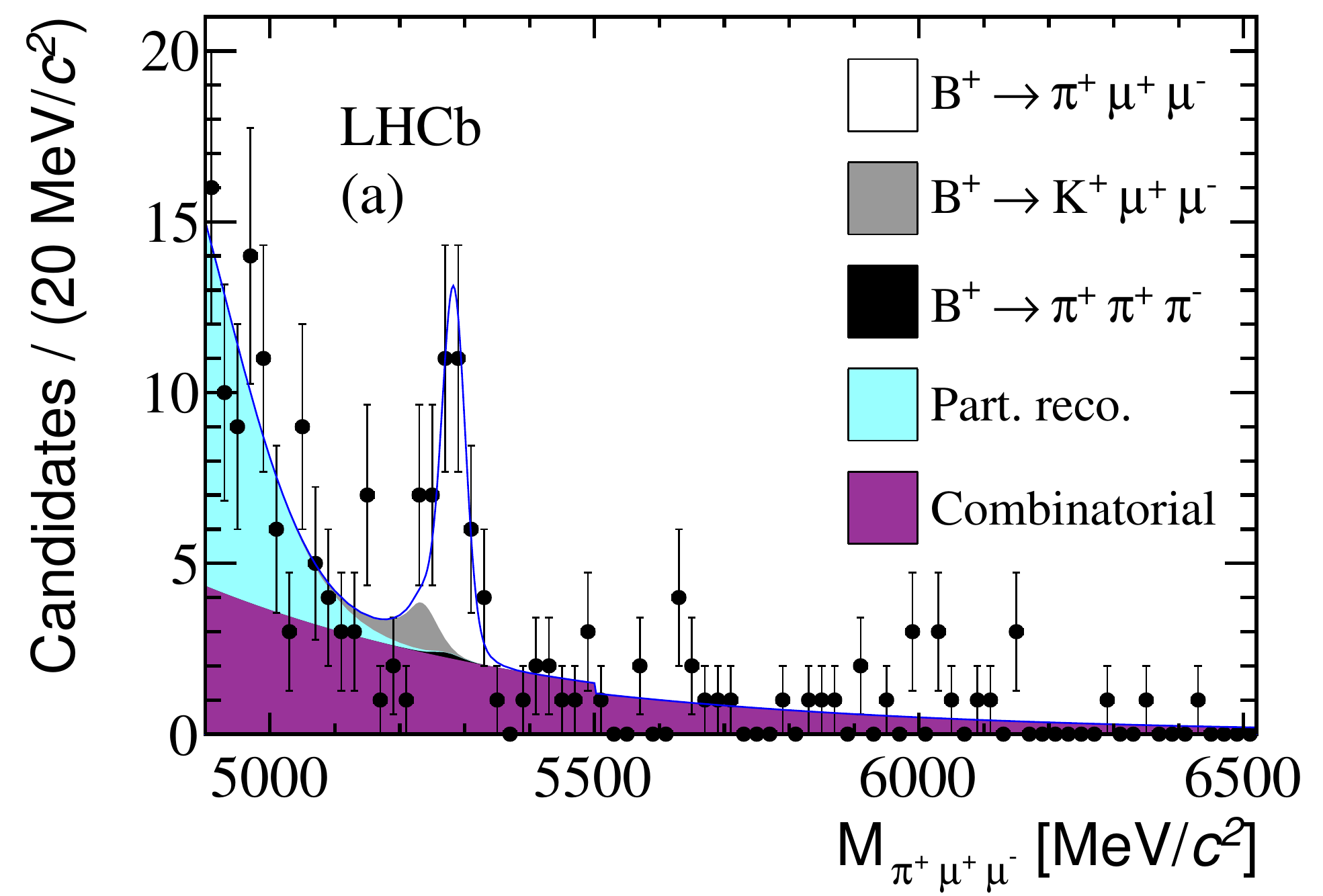}
    \includegraphics*[width=0.33\textwidth]{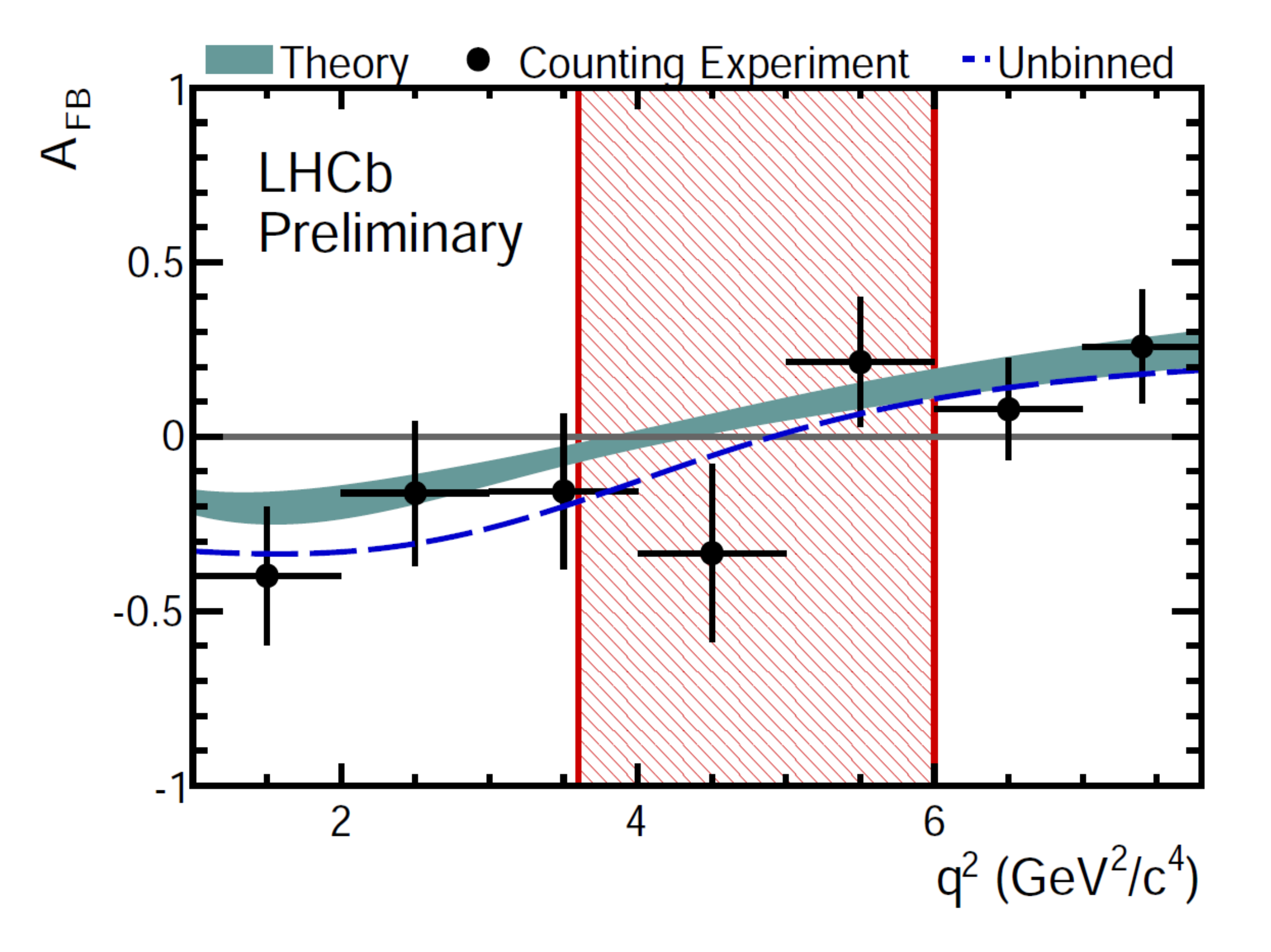}
\caption{The $\pip\mumu$ invariant mass of selected $B^{+} \to \pi^{+} \mu^+ \mu^-$ candidates in 1.0\invfb of integrated luminosity. In the legend, ``part. reco.'' and ``combinatorial'' refer to partially reconstructed and combinatorial background respectively (left). Zero-crossing point of AFB for $B^{0} \to K^{*0} \mu^+ \mu^-$ decays (right).}
\label{fig:angular}
\end{center}
\end{figure}

The LHCb measurements of these quantities~\cite{angularlhcb} show no deviation from the SM predictions, and are the most precise ones to date. All are compatible with the SM, but there is still room for NP.
At the point where A$_{\rm FB}$ changes sign, known as zero-crossing point ($q^{2}_{0}$),  the dominant theoretical errors due to form factors calculations cancel out. This point is predicted to lie in the range [$4.0 < q^{2}_{0} < 4.3$\,GeV$^2$/c$^4$]~\cite{crossing}. LHCb has measured for the first time the zero-crossing point $q^{2}_{0} = 4.9^{+1.1}_{-1.3}$\,GeV$^2$/c$^4$, in agreement with the SM predictions (Figure~\ref{fig:angular}).

\section{Isospin asymmetry of $B \to K^{(*)} \mu^+ \mu^-$}\label{isospin}

LHCb has been able to select 60 \decay{\Bd}{K^{0}\mumu} decays, where
\decay{\KS}{\pip\pim}, reporting an observation at 5.7$\sigma$ for this decay, and 80 $B \to K^{(*)} \mu^+ \mu^-$ decays, where 
\decay{K^{*+}}{\KS\pip}, which are comparable in size to 
the samples  available for these modes in the full data
sets of the $B$-factories. The isospin asymmetry of these decays  is defined as:
\begin{equation}
A_I \equiv \frac{ \mathcal{B}(B^{0} \rightarrow K^{(*)0} \mu^+ \mu^-) - 
  \frac{\tau_0}{\tau_{\pm}}\mathcal{B}(B^{\pm} \rightarrow K^{(*)\pm} \mu^+ \mu^-)}
  { \mathcal{B}(B^{0} \rightarrow K^{(*)0} \mu^+ \mu^-) + 
  \frac{\tau_0}{\tau_{\pm}} \mathcal{B}(B^{\pm} \rightarrow K^{(*)\pm} \mu^+ \mu^-)},
\end{equation}
where $\frac{\tau_0}{\tau_{\pm}}$ is the ratio of the lifetimes of the $B^{0}$ and $B^{\pm}$ mesons.

At leading order, isospin asymmetries  are expected to be zero in the SM. Isospin breaking effects are
sub-leading $\Lambda/m_b$ effects, which are difficult to estimate due
to unknown power corrections. Nevertheless such effects
are expected to be small and these observables may be useful in NP
searches as they offer complementary information on specific
Wilson coefficients~\cite{Feldmann:2002iw}.

\begin{figure}[htb]
\begin{center}
    \includegraphics*[width=0.33\textwidth]{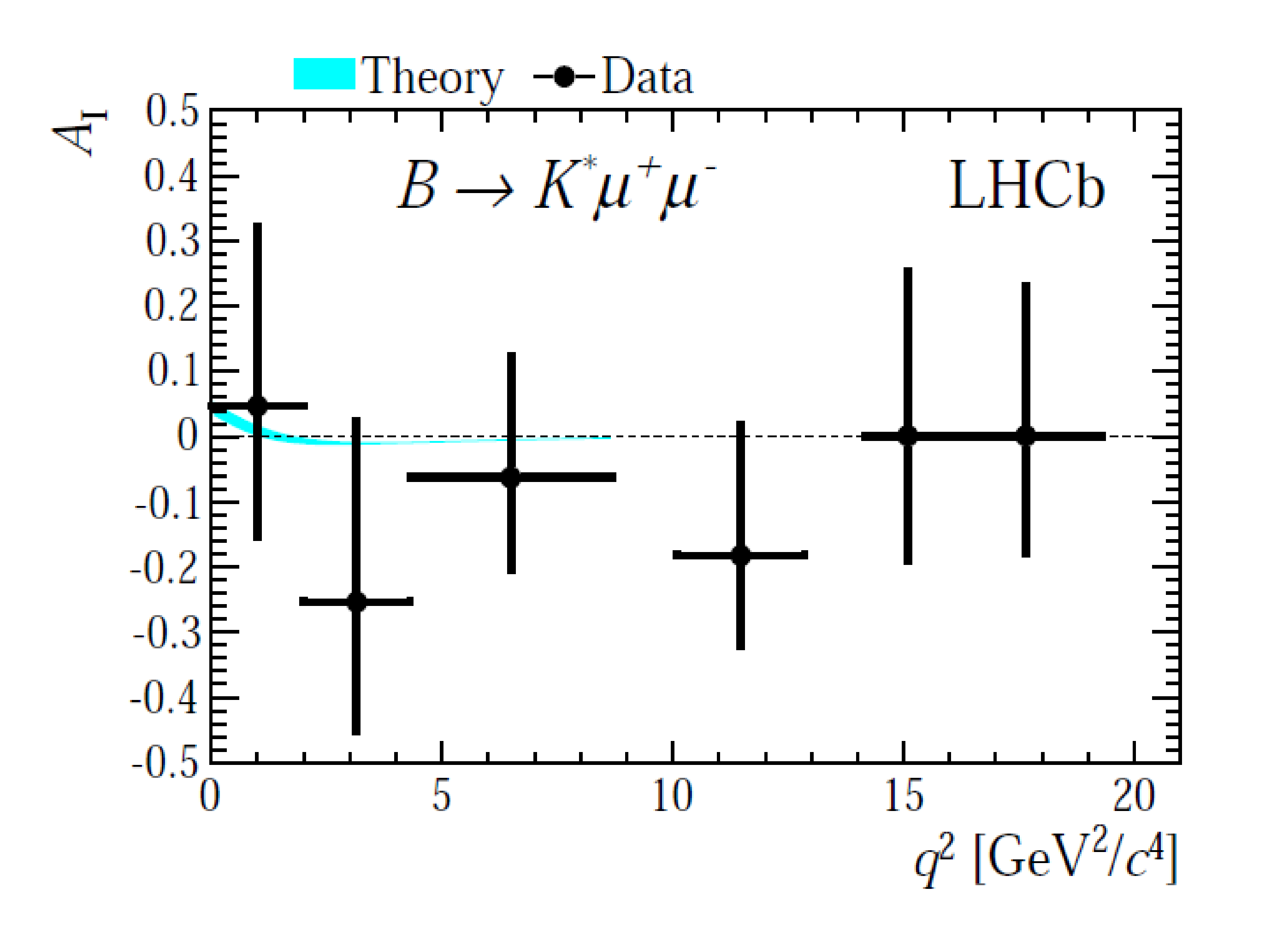}
    \includegraphics*[width=0.33\textwidth]{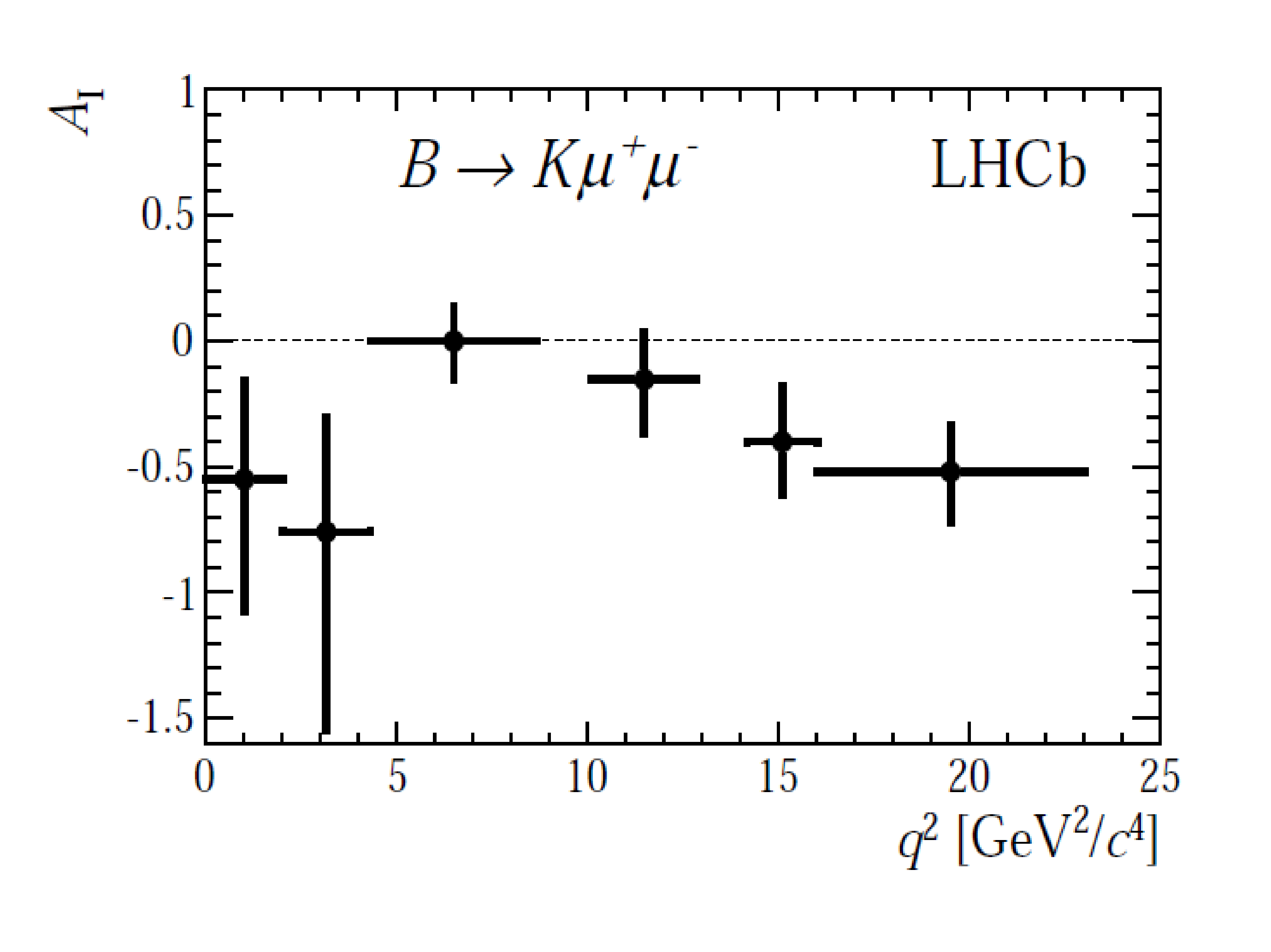}
  \caption{
Isospin asymmetry as a function of $q^2$ in the $B \to K^{*} \mu^+ \mu^-$ (left) and $B \to K \mu^+ \mu^-$ systems (right).}
\label{fig:isospin}
\end{center}
\end{figure}

 The pre-LHC status was an overall consistency with the SM, though BaBar has reported a 3.9 sigma difference with respect to the SM expectation at $q^2$ below the $J/\psi$ mass.
The LHCb measurement of the $K$ and $K^{*}$ isospin asymmetries in
bins of $q^2$ are shown in Fig.~\ref{fig:isospin}.  For the \Kstar
modes $A_{\rm I}$ is compatible with the SM expectation that $A_{\rm
  I}^{SM} \simeq 0$, but for the $\Kp/K^{0}$ modes, $A_{\rm I}$ is
seen to be negative at low- and high-$q^{2}$~\cite{LHCb-PAPER-2012-011}. The two $q^{2}$ bins below 4.3 \gevcc and the highest bin above
16 \gevcc have the most negative isospin asymmetry. These
regions are furthest from the charmonium regions and therefore are theoretically well predicted. 
This is consistent with what seen at previous experiments, but inconsistent with the naive expectation of $A_{\rm I} \sim 0$ at the 4.4 sigma level.

\section{Radiative Decays}\label{radiative}

In the SM, $B$ radiative decays proceed at leading order via electromagnetic 
penguins transitions, while extensions of the SM predict additional one-loop 
contributions.
The direct \CP-asymmetry $A_{\CP}$ in \decay{\Bd}{\Kstarz\gamma} is a powerful observable. BaBar's previous measurement is consistent with the SM expectations.

\lhcb observes $5279\pm93$
\decay{\Bd}{\Kstarz\gamma} and $691\pm36$
\decay{\Bs}{\phi\gamma} candidates, 
respectively~\cite{LHCb-PAPER-2012-019}.
These are the largest samples of rare radiative \Bd\ and \Bs\ decays collected by a single experiment. 
The large sample of \decay{\Bd}{\Kstarz\gamma} decays has enabled \lhcb to make the world's most precise measurement to date of the direct \CP-asymmetry $A_{\CP}(\Kstar\gamma) = 0.8 \pm 1.7 \pm 0.9 \,\%$~\cite{LHCb-PAPER-2012-019}.

This value is compatible with the SM expectation $A_{\CP}(\Kstar\gamma) = -0.0061\pm 0.0043$ \cite{Keum}.
With more statistics, LHCb can perform more analyses of radiative decays, to impose additional constraints on the $C_7 - C_7^\prime$ plane through measurements of $b \to s\gamma$ processes. 
This includes a time-dependent analysis of $\decay{\Bs}{\phi\gamma}$, and  measurements of the photon polarisation through the decays $\Lb \to \Lambda^{(*)}\gamma$.

\section{Conclusions}\label{conclusions}

Rare decays are powerful ways to search for NP beyond the SM. LHCb's great performances position it very well in this field .
Measurements of branching fractions of beauty, charm, and tau decays, as well as of angular distribution of $B^{0} \to K^{*0} \mu^+ \mu^-$ and of the isospin asymmetry of the $B \to K^{(*)} \mu^+ \mu^-$ system have been performed.
All these measurements are of an unprecedented accuracy and are consistent with the SM predictions, except for the measured isospin asymmetry of the $B \to K \mu^+ \mu^-$ system, which shows a 4.4\,$\sigma$ deviation from zero.

\end{document}